# What Does ChatGPT Make of Historical Stock Returns? Extrapolation and Miscalibration in LLM Stock Return Forecasts

Shuaiyu Chen, T. Clifton Green, Huseyin Gulen, and Dexin Zhou[*]

September 2024

## Abstract

We examine how large language models (LLMs) interpret historical stock returns and compare their forecasts with estimates from a crowd-sourced platform for ranking stocks. While stock returns exhibit short-term reversals, LLM forecasts over-extrapolate, placing excessive weight on recent performance similar to humans. LLM forecasts appear optimistic relative to historical and future realized returns. When prompted for 80% confidence interval predictions, LLM responses are better calibrated than survey evidence but are pessimistic about outliers, leading to skewed forecast distributions. The findings suggest LLMs manifest common behavioral biases when forecasting expected returns but are better at gauging risks than humans.

[*]Chen and Gulen are from the Mitchel E. Daniels Jr. School of Business, Purdue University, chen4144@purdue.edu; hgulen@purdue.edu. Green is from Goizueta Business School, Emory University, clifton.green@emory.edu. Zhou is from Zicklin School of Business, Baruch College, dexin.zhou@baruch.cuny.edu. We thank seminar participants at Syracuse University for comments.

## 1. Introduction

Generative artificial intelligence (AI) has shown immense potential in various fields such as transportation, medicine, and economics. Along with the prospect of self-driving cars and improved disease detection, AI holds the potential to transform financial decision-making by objectively analyzing large quantities of information. For example, recent technological advances have been shown to improve the performance of individual investors (Reher and Sokolinski, 2024), sell-side analysts (Cao et al., 2024), and firm auditors (Fedyk et al., 2022). On the other hand, large language models (LLMs) and other AI algorithms are often trained on human output, and research suggests that these approaches may embed harmful social biases (e.g., Gallegos et al., 2024).[1]

To the extent that AI algorithms mimic human decision-making, they may also incorporate cognitive biases that individuals exhibit in financial contexts. An extensive literature documents that people excessively extrapolate from recent performance, tend to be overly optimistic, and are unreasonably confident in their predictions.[2] In this study, we assess the extent to which state-of-the-art generative AI models, as proxied by OpenAI's large language model ChatGPT-4, manifest behavioral biases when provided with historical return data and prompted for stock return forecasts.

While "Large Language Model" might suggest a primary focus on textual data, these models possess capabilities that extend well beyond language processing. For instance, ChatGPT can interpret and analyze visual elements such as plots and charts, demonstrating skill in handling visual and numeric data. We begin by examining how LLMs interpret the timing of historical return data. Several studies have found evidence that investors' expectations about an asset's future return are a positive function of recent past returns, with excessive weights on recent versus distant return realizations (e.g., Greenwood and Shleifer, 2014; Barberis et al., 2015, Cassella and Gulen, 2018; Kuchler and Zafar, 2019; Atmaz et al., 2023).

---

[1] For example, the introduction of machine learning has been shown to disproportionately favor white borrowers in credit screening applications (Bartlett et al., 2022; Fuster et al., 2022; Bowen et al., 2024).
[2] Hirshleifer (2015) and Barberis (2018) provide reviews.

In one study, Da, Huang, and Jin (2021) (DHJ) examine investors' individual stock return expectations using data from Forcerank, a unique crowdsourcing platform for ranking stocks. In each contest, participants rank ten stocks based on their perceived future performance over the following week, providing precise ranking data with a clear forecasting horizon for a set group of stocks. In our first AI investigation, we query ChatGPT to "compete" in each of the 1,379 stock ranking contests while providing 12 weeks of historical return data for the ten stocks in each contest.

Consistent with DHJ, we confirm that human performance rank forecasts place positive weights on all 12 historical returns, with the greatest positive emphasis on the previous week and each preceding week becoming generally less important. In stark contrast to human predictions, realized return regressions for stocks used in these contests show evidence of weekly return reversals, with negative coefficients on the past four return lags, including one significant coefficient.

It is unclear what relation will emerge between ChatGPT forecasted returns and historical return data. While humans are prone to over-extrapolation, the phenomenon of short-term return reversals has been well-established for decades (Jegadeesh, 1990; Lehmann, 1990). For example, Jegadeesh (1990) documents significant profits using a reversal strategy based on prior month returns, and this work has received close to 4,000 Google Scholar citations. It is possible that ChatGPT's training may incorporate stylized facts about short-term stock reversals. In addition, our query provides exclusively numeric data rather than human-authored text, which may place the algorithm in a less behavioral, more mathematical context. Although LLMs are not specifically designed to handle numeric tasks, their ability to recognize patterns, learn from statistical correlations, and understand context allows them to effectively approximate numeric operations.[3]

Empirically, the observed correlation between the average human stock-level forecast rank (a number between 1 and 10) and the matching GPT4 forecast rank is 0.279, which suggests a significant commonality between the two forecasts. The regression evidence reveals that both humans and GPT4 forecasts rely on past data in surprisingly similar ways. As with human forecasts, the coefficients on lagged returns are positive, with the first lag being the largest, the second lag the next largest, the third lag the next largest, and the remaining coefficients noticeably smaller. In light of the empirically observed short-term return reversals, LLM's extrapolation is counterproductive and produces rankings that are negatively associated with future performance.

Participants in the Forcerank contests extrapolate negative and positive returns asymmetrically, consistent with neuroscience studies that show gains and losses are processed by different regions of the brain (Kuhnen and Knutson, 2005). Specifically, humans tend to place greater emphasis on negative returns, and negative performance has a longer-lasting effect on expectations (Gulen and Lim, 2024). In contrast, we find evidence that GPT4 forecasts place greater emphasis on recent positive returns than negative returns, while extrapolating more distant negative returns in ways similar to humans.

We next shed light on how LLMs interpret visual financial data. Specifically, we create 12- and 24-week price charts for each contest stock illustrating the open, high, low, and close prices for each day (e.g., Jiang, Kelly, and Xiu, 2023). The resulting price charts for each contest stock are included in the query, and ChatGPT-4 is prompted to issue performance rank forecasts for the following week. In line with the prompts that include numerical return data, ChatGPT's forecasts continue to extrapolate from past returns when visually gathering return information from the price charts.

---
[3] Although LLMs struggle with complex math (e.g., Lohr, 2024), ChatGPT-4 can approximate the mean and standard deviation of a series of data when prompted. Van and Cunningham (2024) find evidence that ChatGPT-4 can predict Oscar winners and economic trends post-training.

The Forcerank setting emphasizes cross-sectional variation in return performance. We also explore how ChatGPT forms aggregate market return forecasts using rolling windows of monthly historical return data for the S&P500 Index. We compare LLM aggregate sentiment characteristics to investor expectations inferred from survey data from the American Association of Individual Investors (as in Greenwood and Shleifer, 2014). Consistent with human expectations, we find that GPT4 return forecasts place the largest positive weights on recent returns.

We next consider sentiment forecasts from an alternative LLM. In particular, we repeat the S&P500 return-based queries using the Claude large language model from Anthropic. Although Claude and ChatGPT were trained independently, the correlation between the two sentiment measures is 0.78, suggesting that the same underlying human behavior is manifested in their responses. The extrapolative coefficients from the Claude forecasts are very similar to the estimates from ChatGPT, suggesting that return extrapolation in LLM forecasts is not confined to individual stocks or a single LLM.

In our next set of queries, we investigate how LLMs react to the relative magnitudes of individual historical returns when predicting distinct characteristics of the return distribution. Ben-David, Graham, and Harvey (2013) (BGH) survey CFOs regarding their projections of the overall market and their individual companies. A unique survey feature is that executives were asked to predict 80% forecast confidence intervals. The study documents significant miscalibration, with realized returns falling outside projected intervals substantially more than 20% of the time. In order to evaluate the calibration of LLM stock return forecasts, we randomly select 10,000 stock-month observations. For this set, we gather ten years of historical monthly return observations and prompt ChatGPT to answer questions similar to the survey in BGH.

We first examine whether LLM expected return forecasts appear biased relative to realized outcomes. Humans tend to be overly optimistic in a variety of settings (e.g., Van den Steen, 2004). If overoptimism manifests in the training data, then LLM forecasts may also be higher than realized returns. Empirically, we find consistent evidence that LLM expected returns are significantly higher than both historical means and realized returns. For example, the average monthly cross-sectional historical mean provided to ChatGPT is 1.4%, and the cross-sectional average for the next period's realized return is 1.15%. Yet the average expected return projected by GPT4 is roughly twice as large at 2.2%. The observed ChatGPT positive bias is partially attributable to return forecasts being largely truncated at zero, suggesting that the training data may have embedded the economic idea that expected returns should be nonnegative.

Turning to low and high forecasts, we observe that next month's realized return value lies within the GPT4 80% confidence interval 76.9% of the time, which is less accurate than the 79.0% that could be contained by simply using the $10^{th}$ and $90^{th}$ historical percentiles as the forecasts. However, the evidence of miscalibration in ChatGPT forecasts is much less severe than in the executive surveys, consistent with enhanced numeracy for LLMs. Nevertheless, we observe that the $10^{th}$ percentile LLM forecast (Low) is significantly less than the $10^{th}$ historical percentile, suggesting pessimism in projecting unfavorable outcomes. On the other hand, the $90^{th}$ percentile forecast (High) is also significantly less than the $90^{th}$ historical percentile.

To deepen our understanding of how LLMs translate historical returns into forecasts, we regress LLM return forecasts on historical $10^{th}$, $20^{th}$, …, $90^{th}$ percentiles. Unsurprisingly, when the dependent variable is the forecast of next period's return, the regression produces significant loadings on each of the nine percentile measures. However, we observe that the largest loading is on the 90th percentile, consistent with forecast overoptimism. Examining how Low and High forecasts incorporate historical data, we observe that both forecasts load significantly on the corresponding percentiles, but they also load significantly on percentiles on the other side of the distribution with negative signs, indicating underlying assumptions about distributional symmetry. However, the High forecast is less sensitive to high percentiles than the Low forecast is to low percentiles, suggesting underlying pessimism about the tails of the distribution.

Taken together, the analysis demonstrates that LLM stock return forecasts exhibit over-extrapolation of historical return performance. While LLM forecasts are considerably better calibrated than human forecasts, indicating improved assessments of risk, LLM forecasts tend to be excessively optimistic when predicting expected performance and slightly pessimistic about the tails of the distribution. As a result, LLM forecasted return distributions are positively skewed compared to historical data.

The findings contribute to recent literature that examines the extent to which LLMs reproduce human behavior in financial contexts.[4] For example, Horton (2023) shows that LLMs respond to standard economic experiments in ways similar to humans, and Fedyk et al. (2024) finds evidence that ChatGPT embeds investment preferences that vary across gender, income, and age. Evidence on using LLMs to generate return forecasts is mixed. Lopez-Lira and Tang (2023) finds that ChatGPT can successfully forecast daily stock returns using news headlines, and Kim, Muhn, and Nikolaev (2024) find ChatGPT excels at distilling corporate disclosures, suggesting LLMs may outperform humans at interpreting news. On the other hand, Bybee (2023) infers LLM expectations from newspaper articles over longer horizons and finds evidence of human-like extrapolative sentiment.

---

[4] LLMs have also been shown to produce realistic human responses in marketing and political science contexts (e.g., Li et al., 2023; Argyle et al., 2023).

Our analysis innovates by exploring how LLMs build forecasts using numeric data, and we are able to closely analyze how inputs translate into forecasts for both humans and LLMs in a similar context. Although we find evidence of successful risk assessments, the findings generally caution against assuming that LLMs interpret even straightforward numeric data with fully rational statistical rigor. More generally, our study contributes to the broader discourse on the integration of AI in financial decision-making and the critical need to address potential biases.

## 2. Data Collection: Investor and LLM Stock Return Forecasts

In this section, we describe the samples of human stock performance forecasts and the methodology for collecting the analogous ChatGPT-generated forecasts.

Our first source of human forecast data is from Forcerank, a crowd-sourced platform for ranking stocks that is hosted by Estimize. Forcerank organizes weekly competitions in which participants rank a list of ten stocks according to their perceived return performance (percentage gain) over the next week. Participants' goal is to rank the ten stocks from one to ten based on their perception of the stocks' rankings according to next week's realized returns. Higher performance ranks receive higher scores. Forcerank assigns points to participants based on the accuracy of their rankings and maintains weekly leader boards that reflect cumulative performance (see Da, Huang, and Jin, 2022 for more details).[5]

The sample contains 1,283 weekly contests including a total of 200 unique stock tickers. As in DHJ, we use each contest stock's average score that ends in week $t$ as a proxy for investors' consensus expectations at time *t* about stock returns over week *t* + 1. We focus on contests that refer to the prediction of future returns and contest categories outlined in DHJ. We ensure that consensus expectations are regressed on returns that investors have observed prior to submitting their ranking to Forcerank. To this end, we measure consensus expectations based on forecasts submitted to Forcerank only by those investors who observe stock returns ending in week *t*. All contests in our analysis begin on Monday morning of week *t* + 1, and we use calendar trading-week returns and performance ranks in weeks prior to *t* as the primary independent variables of interest.

---

[5] Forcerank initially offered cash prizes, but the SEC considered the practice to be an illegal security-based swap (https://www.sec.gov/newsroom/press-releases/2016-216). Dropping this feature reduced interest and Forcerank was shut down in 2018. Cassella et al. (2023) also studies Forcerank data.

Our goal is to compare LLM forecasts to similar ranks submitted by humans. DHJ examine extrapolative behavior by analyzing how average Forcerank scores load on twelve weeks of lagged stock returns. In our main analysis, we similarly consider twelve weeks of lagged stock returns for each contest stock. We create .csv files for each contest that contain a 10 by 12 grid of weekly stock returns and provide the following prompt to GPT4:[6]

> The following is the return data for ten stocks from week t-12 to week t-1:\n\n
> Based on the information, please rank the return of these ten stocks in week t. How confident are you about the ranking?
> Your output will be in JSON format with the following format:
> '{"rank":{"1":"stock id","2":"stock id",...,"10":"stock id"}, "confidence": }'. 1 stands for the highest return and 10 for the lowest returns.[7] Confidence represents a probability that ranges from 0 to 1.[8]

An important concern with LLM forecasts is that they may be subject to look ahead bias, in which the training data may include future outcomes that can influence forecasts (e.g., Glasserman and

---

[6] More specifically, we use the GPT-4o endpoint for our analyses. It has been shown to be one of the most capable LLMs available at the time of the analysis. See https://openai.com/index/hello-gpt-4o/
[7] Our prompt follows the 1-is-best approach of the Forcerank contests, but as in DHJ we reorder to a 10-is-best rank measure that is more intuitive in the context of the forecast analysis.
[8] ChatGPT's average forecast confidence level is 0.73 with a standard deviation is 0.15. We find no evidence that adjusting forecasts for the level of confidence improves forecast accuracy or changes inferences regarding extrapolation or miscalibration.

Lin, 2023; Sarkar and Vafa, 2024). We follow the recommended strategy of anonymizing the prompts by providing only numeric data for each stock with no firm identifying information.

The Forcerank setting emphasizes stock-level cross-sectional performance. We also consider forecasts of aggregate market performance. To gauge human expectations, we obtain data from the American Association of Individual Investors (AAII) Investor Sentiment Survey. The AAII survey is a weekly survey of the AAII members running from 1987 up to the present day which measures the percentage of participants that are bullish, bearish, or neutral on the stock market for the next six months. We follow Greenwood and Shleifer (2014) and measure expectations using the difference in the bull and bear percentages at the monthly frequency.

For the LLM forecasts of aggregate market returns, we provide S&P 500 index returns in months $t$-12 to $t$-1 returns in a .csv file and provide the following prompt to approximate the AAII Survey:

> The csv data contain the monthly stock returns in months t-12 to t-1.
> Please answer the following questions:
> Do you feel the direction of the stock market over the next six months will be up (bullish), no change (neutral) or down (bearish)?
> How confident are you about this prediction?
> Your output will be in json format with the following format:
> '{"prediction":,"confidence":}'. 1 stands for bullish, 0 for neutral and -1 for bearish.
> Confidence represents a probability that ranges from 0 to 1.

The resulting ChatGPT market sentiment measure is -1 for bearish, 0, for neutral, and 1 for bullish.

Large language models can interpret images as well as numerical data, and we also examine how LLMs translate price charts into performance forecasts. For each contest stock, we create a candlestick price chart that plots the open, high, low and close for each day after normalizing the beginning stock price to $100. An example of one set of contest stock price charts is displayed in Figure 1. Days in which the close was higher than the open are colored green, and days with

negative open-to-close returns are colored red. We then submit the following image-based queries to ChatGPT-4:

> The charts contain daily stock price data for ten stocks from the past 12 weeks.
> The file names of the images contain the stock id.
> Based on the information, please rank the returns of these ten stocks in the following week.
> Your output will be in json format with the following format:
> '{"1":"stock id","2":"stock id",...,"10":"stock id"}'. 1 stands for the highest return and 10

for the lowest return.

Prior analyses largely focus on the direction or ranking of future stock returns. While these prompts are helpful in understanding the expectation of future returns, they are not informative about LLMs' prediction of the future return distributions. Thus, in addition to examining expected returns, we also seek to understand how LLMs determine other aspects of the forecasted return distribution. We are guided by the CFO survey examined in Ben-David, Graham, and Harvey (2013). The Duke CFO has been surveying executives since 1996 and along with overall outlook questions, the survey includes additional questions that can vary over time (e.g., questions related to upcoming presidential elections). BGH examine survey vintages that ask executives to forecast actual aggregate market returns as well as providing the 80% forecast confidence intervals. Motivated by BGH, we prompt ChatGPT4 to issue $10^{th}$ and $90^{th}$ percentile forecasts in addition to expected returns.

We deviate from the executive survey by asking ChatGPT4 to produce stock-level return forecasts instead of aggregate market forecasts. Stock-level data offers greater variation in the historical return distributions, and it also permits many more observations than a single aggregate market return series can provide. In our approach, we randomly select 100 months from the 1926 to 2023 time period. For each selected month, we choose 100 stocks with 10 from each size decile.

For this set of 10,000 stock-months, we gather up to ten years of historical monthly returns (requiring no fewer than five years of returns). We place each set of observations in a .csv file and prompt GPT4 with the following text guided by the survey questions in BGH.

> Below are the monthly returns for a financial asset over the past 120 months.
> Please answer the following questions on next month's return
> There is a 1-in-10 chance the actual return will be less than a%.
> I expect the next month's return to be: b%.
> There is a 1-in-10 chance the actual return will be greater than c%.
> Please return a JSON object in the following format:
> '{"low": a%,"expected": b%,"high": c%}'.

With these samples, we investigate the process by which LLMs translate lagged return data into forecasts, and we compare them with human forecasts and realized outcomes.

Table 1 presents descriptive statistics. Our sample contains 1283 Forcerank contests. Requiring historical return data from CRSP and firm information from Compustat results in a sample of 12,807 stock-contest observations. For this sample, 200 unique stocks are represented. The Forcerank contests attracted 1,757 unique participants, and on average 12 individuals competed in each contest.

The American Association of Individual Investor survey sample covers July 1987 through June 2024 and is comprised of 438 observations. We observe that the average surveyed bull - bear spread is 6.5%, indicating that 6.5% more respondents were bullish about the stock market over the next six months than bearish. The ChatGPT Sentiment score in the sample, which is -1 if bearish, 0 if neutral, and 1 if positive, is 0.37.

### 3. Large Language Model Expectations Formation

Large Language Models are built using deep learning, a technique modeled on the human brain in which a software network of billions of neurons is exposed to trillions of text string training data examples to discover inherent patterns. Instead of associating specific words with individual neurons inside an LLM, words or concepts are associated with the activation of complex patterns of neurons. Since LLMs are essentially grown by training on text strings rather than being explicitly programmed, it makes them black boxes, and research is necessary to uncover how LLMs make decisions.

Although LLMs are not explicitly programmed for numerical tasks, they demonstrate surprising proficiency by recognizing and replicating patterns within the data. They encounter many numerical relationships and operations during training, which fosters a form of statistical learning that allows them to approximate numeric functions by identifying correlations. In addition, LLMs use contextual understanding to apply logic and reasoning that often mirrors mathematical processes. This enhances their ability to perform tasks such as estimation, comparison, and basic arithmetic. On the other hand, mathematical expressions often rely on assumptions and unmentioned rules, and LLMs' reliance on statistical patterns can lead to incorrect responses (e.g., Satpute et al., 2024). Moreover, training on human textual discussions of financial data may introduce behavioral biases into LLM's numeric responses.

Researchers have attempted to reverse engineer the inner workings of LLMs using autoencoders (essentially smaller neural networks) to analyze when small groups of neurons fire together, creating mind-maps that reveal a set of the "features" the LLM has learned (e.g., Bereska and Gavves, 2024). In our analysis, we seek to understand how LLMs interpret the timing and magnitude of historical stock returns when generating return forecasts. Additionally, we consider the extent to which low (10th percentile) and high (90th percentile) forecasts represent distinct LLM features that potentially weigh returns differently.

*3.1 Historical Return Timing – Extrapolation*

In this section, we examine how human and LLM forecasts interpret the timing of historical returns when generating performance forecasts. We consider two settings. First, we ask GPT4 to replicate the Forcerank contest environment, in which humans are asked to predict relative performance for a sample of ten stocks over the following week. We next consider survey evidence from the Yale Center for Finance and the American Association of Individual Investors regarding assessments of aggregate market performance over horizons from one month to one year. Our emphasis is on the extent to which LLMs extrapolate from recent returns in ways similar to humans.

*3.1.1 Performance Rank Analysis*

We begin by analyzing how the timing of lagged returns influences forecasted and realized return performance using the following regression:

$$Y_{i,t} = \gamma_0 + \sum_{s=0}^{n} \beta_s \cdot \text{Return}_{i,t-s} + \varepsilon_{i,t}, \tag{1}$$

where $Y_{i,t}$ is the human Forcerank score for stock *i* in week *t*, the ChatGPT-4 forecasted rank for the same contest-stock, or the realized performance for the stock in week *t*+1. $\text{Return}_{i,t-s}$ represents lagged weekly stock returns. We consider 12 and 24 weekly lags as in DHJ. Standard errors are clustered at the contest level.

The results are presented in Table 2. Specification (1) confirms that human forecasts of future performance are strongly influenced by past returns. The coefficients on the past 12 weekly returns are all positive and mostly significant, with the magnitudes being similar to the coefficients in DHJ. Most notably, the coefficients on recent past returns are in general higher than those on distant past returns.

In Specification (2), we examine the evidence for ChatGPT4 forecast ranks. The coefficients show even greater evidence of over-reliance on most recent returns in LLM forecasts. Specifically, the positive coefficient on the previous week is more than 10 times larger than the coefficient for two weeks prior (compared to 3 times larger for humans), and the coefficients continue to decline in previous weeks. Table IA1 in the Internet Appendix considers contest-adjusted returns and also finds evidence of return extrapolation with a strong emphasis on recent periods. Understandably, the R-squares are considerably lower for humans than for ChatGPT (3.4% vs 35.2%) since humans had other information at their disposal at the time of the contest whereas the LLM was only provided with historical returns. However, the coefficients provide clear evidence that past returns drive human and LLM forecasts of future performance in similar ways.

Specification (3) shows that human and ChatGPT rankings are significantly related, and Specification (4) indicates that the relation holds after controlling the lagged returns, suggesting that human and GPT4 rely on returns in ways that are not fully captured by the linear extrapolation model. Specification (5) provides the benchmark by setting the dependent variable to next week's realized return. Consistent with the well-established literature on short-term return reversals,[9] and in direct contrast to human and GPT4 expectations, realized weekly returns exhibit short-term reversals. Many of the coefficients are negative, and the lags at $t$-1, $t$-2, and $t$-2 are statistically significant. The evidence in Table 2 suggests that LLM's training serves to incorporate humans' counterproductive tendency to assume that recent stock return performance will continue.[10]

---

[9] Examples across the decades include Jegadeesh (1990), Lehmann (1990), Avramov, Chordia, and Goyal (2006), Da, Liu, and Schaumburg (2014), and Chui, Subrahmanyam, and Titman (2022).

[10] In Table IA1 in the Internet Appendix, we also consider contest-adjusted returns (i.e., the stock return in excess of the average return of the ten stocks in the contest) and find similar (stronger) evidence of extrapolation in LLM forecasts.

In Table IA2 in the Internet Appendix, we explore whether providing more data reduces the emphasis on recent returns. In particular, we expand from 12 to 24 weeks of lagged returns in the data provided in the GPT4 prompt. Although including the additional 12 lags results in a few additional significantly positive coefficients for both humans and GPT4, the coefficient magnitudes continue to place strong emphasis on the most recent weeks.

Humans have been shown to react asymmetrically to gains and losses (e.g., Kuhnen, 2015), and neuroscience studies show that positive and negative return extrapolations activate different regions of the brain (Kuhnen and Knutson, 2005). We next examine whether LLM forecasts also embed asymmetric reactions to historical returns. In particular, we decompose lagged returns into two separate measures using the following model.

$$\text{Forecast Rank}_{i,t} = \gamma_0 + \sum_{s=0}^{n} \beta_s^+ \cdot \text{Return}_{i,t-s}^+ + \sum_{s=0}^{n} \beta_s^- \cdot Return_{i,t-s}^- + \varepsilon_{i,t}, \quad (2)$$

where $\text{Return}_{i,t-s}^+$ is equal to the return for stock $i$ in week $t$ if it is positive, otherwise zero, and similarly $\text{Return}_{i,t-s}^-$ is the weekly return if negative, else zero. Forecast Rank$_{i,t}$ is either the Forcerank score or the ChatGPT-4 forecast rank. The results are presented in Table 3. As in DHJ, we observe that humans react much more strongly to negative performance than positive performance, and the weights decay more slowly into the past. Only one lag of positive returns in Specification (1) is significant, whereas all twelve negative return lags are significant. Moreover, the magnitude of the first negative lag coefficient is almost twice as large as the coefficient on the first positive lag.

In contrast to human behavior, we observe in Specification (2) that the strongest extrapolation in ChatGPT forecasts occurs for recent positive returns. The coefficients on one- and two-week lags are larger for positive returns for negative lags. However, we do observe that LLM forecasts exhibit the same tendency to react more strongly to distant negative returns as with human forecasts, with coefficients on negative returns remaining significant at longer lags. In sum, LLM forecasts appear more symmetric than human forecasts but continue to emphasize distant negative returns in human ways.

Forcerank contests are geared toward predicting relative performance, so we also consider variants of Equation (1) where we set the dependent variable equal to the realized performance rank for the contest stock, and we also consider historical return ranks as the independent variables. Table 4 presents the results. In Specification (5), we continue to observe evidence of short-term reversals in realized performance when returns are ranked among a set of ten stocks. In contrast, both human and ChatGPT forecasts show strong positive extrapolation of past performance rank. While Forcerank scores load on several lags, GPT4 performance ranks are only significant for two lags, with last week's performance rank weighing heavily and helping lead to an R-squared of 0.82.

Large language models can interpret images as well as numerical data, and we next examine how ChatGPT translates price charts into performance forecasts. Figure 1 depicts examples of price chart images that we provide to ChatGPT-4 while prompting for next-week performance ranks. We then repeat the extrapolation regression in Equation (1) using price chart-based performance rank forecasts.

The results for the 12-week price chart forecasts are presented in Table 5. Perhaps unsurprisingly, the magnitudes of the lagged return coefficients are smaller for the candlestick price charts than for the numerical return-based forecasts. However, the coefficients continue to display a near monotonic downward trend in coefficient size over more distant lags. The findings are similar when using 24-week price charts as inputs for the forecasts, as reported in Table IA3.

The price chart evidence suggests that the inclination for LLMs to extrapolate from past returns extends beyond numeric data to unstructured image data.

The results from the linear regressions in Equation (1) indicate a clear and robust decay pattern in the relation between human and LLM performance ranks and recent past returns. To capture this pattern parsimoniously, we next estimate a parametric regression model that assumes an exponential decay of weights on past returns as follows[11]:

$$Y_{i,t} = 5.5 + \lambda_1 \cdot \sum_{s=0}^{n} w_s \text{Return}_{i,t-s} + \epsilon_{i,t}, \qquad (3)$$

where $w_s = \frac{\lambda_2^s}{\sum_{j=0}^{n} \lambda_2^j}$ and $Y_{i,t}$ is either the Forcerank score or the GPT4 forecast rank.

The first parameter ($\lambda_1$) is a scaling factor that multiplies all past returns of stock $i$ and captures the level effect, i.e., the overall extent to which investor expectations respond to past returns. The second parameter ($\lambda_2$) captures the slope effect that governs how past returns are relatively weighted in forming expectations, with a $\lambda_2$ closer to zero meaning that investors put higher weight on recent past returns as opposed to distant past returns. A higher $\lambda_1$ and a lower $\lambda_2$ jointly lead to a higher degree of extrapolation, leading to the degree of extrapolation measure $\lambda_1(1 - \lambda_2)$.

Table 5 presents the regression estimates for the exponential decay model. The level coefficient ($\lambda_1$) is larger for GPT4 forecast ranks than for humans, but this is perhaps to be expected since humans have other sources of information at their disposal to shape their expectations. Moreover, the slope coefficient ($\lambda_2$) is smaller for GPT4, suggesting an even greater weight on recent past returns, and resulting in a higher degree of extrapolation than human forecasts (38.1 vs 12.09 for humans). Results from ChatGPT ranks generated based on 24 lagged weekly returns are similar.

---

[11] In addition to DHJ, Greenwood and Shleifer (2014), Barberis (2015), and Cassella and Gulen (2018) have also used this approach.

The evidence that LLM forecasts extrapolate recent returns, combined with the evidence of short-term return reversals, suggests that GPT4 forecasts may negatively predict future returns. On the other hand, the linear extrapolation model in Table 2 explains only 38% of the variation in GPT4 forecasts, and it is possible that non-extrapolating aspects of LLM forecasts may be positively related to returns.

We explore the relations between forecasts and future returns using Fama-MacBeth regressions, in which the dependent variable is the daily return of an individual stock over the next week. To better understand the source of return predictability, we decompose LLM and human forecasts into two components: a predicted score and the residual. The predicted score is computed as the fitted value from the regression in Eq. (1). In other words, *Predicted GPT4* is the weighted average of the past 12 weekly returns that best predicts the LLM forecast, and the residual of this regression is labeled *Residual GPT4*. Predicted and residual human Forcerank scores are computed analogously. We consider specification with the following set of firm controls that have been shown to forecast future stock returns: log market capitalization, log book-to-market, asset growth, gross profits-to-assets, market beta, weekly turnover, and the max daily return in the last month. All control variables are measured in the month of week *t* and prior to the return in *t*+1.

Table 7 reports these regression results. We find evidence that both human and LLM forecasts are negatively associated with future returns, and the evidence is most robust for Predicted forecasts, indicating that the extrapolative aspect of human and LLM forecasts is most negatively predictive of performance. The return predictability evidence is consistent with the view that training out human output can result in LLM return forecasts that predict future returns with the wrong sign.

*3.1.2 Sentiment Analysis*

The Forcerank setting analyzes the relative performance in a cross section of stocks. In this section, we examine GPT4 expectations measures of aggregate market performance to survey evidence from the American Association of Individual Investor survey.

$$\text{Sentiment}_{i,t} = \gamma_0 + \sum_{s=0}^{n} \beta_s \cdot R_{i,t-s} + \varepsilon_{i,t}. \tag{4}$$

For human sentiment, $\text{Sentiment}_{i,t}$ for stock $i$ in month $t$ is a number that ranges from -1 to 1 and captures the percentage of individual investors that expect the stock market will go up over the next 6 months less the fraction of investors that think the stock market will go down.[12] For ChatGPT, $\text{Sentiment}_{i,t}$ is equal to 1 if the GPT4 assesses based on historical returns, that the direction of the stock market over the next six months will be up, 0 if expecting no change, and -1 if expecting down. $R_{i,t-s}$ denotes S&P 500 index returns. The regression results are presented in Table 8. Consistent with human expectations, we find that ChatGPT return forecasts place larger positive weights on recent returns. In particular, the largest coefficient is on the first return lag, and the coefficients decline monotonically for the next several weeks.

We next consider S&P 500 return aggregate market sentiment measures generated using an alternative large language model. Specifically, we repeat the market sentiment queries using another LLM that is widely regarded at the time of the analysis, the Claude 3.5 Sonnet model from Anthropic. Table 9 presents the results. We observe that the two different LLM sentiment forecasts are closely related. Regressing the Claude sentiment measure on the ChatGPT sentiment produces a coefficient of 0.81 and an R-squared of 0.603, suggesting a correlation of 0.78 between the two sentiment measures. Moreover, the magnitudes of the lagged return coefficients are very similar when comparing Specification (2) of Table 7 to Specification (1) of Table 8. Together, the sentiment analysis evidence helps confirm that return extrapolation in ChatGPT forecasts is not confined to individual stocks or a specific large language model.

---

[12] Neutral is also an option, i.e., bear does not equal (1–bull).

*3.2 Historical Return Magnitudes – Bias and Miscalibration*

In this section, we ask GPT4 to provide specific forecasts about the characteristics of the stock return distribution. Our guiding setting is the company executive survey analyzed in Ben-David, Graham, and Harvey (2013), which asks CFOs to forecast actual returns as well as providing 80% forecast confidence intervals. Motivated by BGH, we provide monthly historical return data and prompt GPT4 to issue next-period return forecasts in addition to $10^{th}$ and $90^{th}$ percentiles for a random sample of 10,000 stock-months.

We first examine whether LLM expected return forecasts appear biased relative to realized outcomes. Humans tend to be overly optimistic in a variety of settings. For example, Van den Steen (2004) argues that investors are more likely to choose stocks for which they have overestimated the likelihood of success, similar to the winner's curse. If overoptimism manifests in the training data, then LLM forecasts may also be higher than realized returns. Table 10 presents forecast statistics for the sample of 9,959 stock-months that survived the historical return requirements. We observe that the cross-sectional mean of the GPT4 forecast for next month's stock return is 2.2%. The LLM forecast is considerably larger than the average historical mean for the data that were provided in the prompt (1.4%) and roughly twice the magnitude of next month's realized return (1.1%), with both differences being statistically significant.

Figure 2 (middle panel) plots the cross-sectional distribution of the historical means of the data provided in the prompts and the distribution of resulting GPT4 forecasts. The distribution of historical means appears Gaussian, suggesting that the historical sample sizes are sufficient for the central limit theorem to apply. On the other hand, GPT4 forecasts are decidedly less smooth. Very few of the GPT4 expected return forecasts are below zero (0.45% of forecasts), which suggests that the LLM's training may have embedded the idea that expected returns should be nonnegative. However, Table 11 indicates that the median, 75$^{th}$, and 90$^{th}$ percentile of the distribution of GPT4 expected return forecasts are all higher than the average historical equivalents, suggesting that ChatGPT's positive expected return bias extends beyond truncating at zero.

We next examine LLM forecasts of Low and High returns. The average GPT4 80% confidence interval is 23.4%, which is smaller than the average historical confidence interval of 25.5%. We observe that next month's realized return value lies within the GPT4 confidence interval 76.9% of the time, which is less accurate than the 79.0% that could be contained by simply using the 10$^{th}$ and 90$^{th}$ historical percentiles as the forecasts. The miscalibration is primarily on the upside, with 12.7% of realized returns occurring above the High forecast, compared to 10.4% on the downside.

The evidence of miscalibration in ChatGPT forecasts is much less severe than in the executive surveys, with BGH reporting that realized market returns are within executives 80% confidence intervals only 36% of the time. The findings of substantially better calibration of GPT4 forecasts relative to humans is consistent with improved numeracy. Despite the evidence of generally being well-calibrated, the GPT4 Low forecast is significantly less than the 10$^{th}$ historical percentile, suggesting conservatism in projecting unfavorable outcomes. On the other hand, the GPT4 High forecast is also significantly less than the 90$^{th}$ historical percentile. These patterns are evident in Figure 3, which plots the distribution of Low – historical 10$^{th}$ percentile, Forecast – historical mean, and High – historical 90$^{th}$ percentile.

To deepen our understanding of how LLMs translate historical returns into forecasts, we regress return ChatGPT-4 forecasts on characteristics of the historical return distribution as follows:

$$\text{Forecast}_{i,t} = \gamma_0 + \beta_1 Ret_i^{min} + \beta_2 Ret_i^{10\%} + \cdots + \beta_{10} Ret_i^{90\%} + \beta_{11} Ret_i^{max} + \varepsilon_{i,t}. \qquad (5)$$

Forecast$_{i,t}$ is either the GPT4 forecast of the next period return, the Low (10$^{th}$ percentile) forecast, or the High (90$^{th}$ percentile) forecast. $Ret_i^{min}$ and $Ret_i^{max}$ are the minimum and maximum of the historical return distribution provided in the prompt, and $Ret_i^{10\%}$ is the 10$^{th}$ percentile of the historical return data, etc. The results are presented in Table 9.

Unsurprisingly, when the dependent variable is the expected return forecast, the regression returns significant loadings on each of the nine percentile measures. However, we observe that the largest loading is on the 90$^{th}$ percentile, providing additional evidence of positive bias in return forecasts. Examining how *Low* and *High* forecasts incorporate historical data, we observe that both forecasts load significantly on the corresponding percentiles, but they also load significantly on percentiles on the other side of the distribution with negative signs, indicating underlying assumptions about distributional symmetry. However, the *High* forecast is less sensitive to high percentiles than the *Low* forecast is to low percentiles, suggesting underlying pessimism about the tails of the distribution. The evidence that Low and High forecasts have different characteristics suggests that the training data has led ChatGPT to treat high and low stock return outcomes as distinct model features.

Taken together, the findings suggest that LLM forecasts are considerably better calibrated than human forecasts, indicating improved assessments of risk. However, LLM forecasts tend to be excessively optimistic when predicting expected performance and slightly pessimistic about the tails of the distribution. As a result, LLM forecasted return distributions are positively skewed compared to historical data.

*3.3 Discussion*

Our analysis suggests that LLM stock performance forecasts exhibit excessive extrapolative behavior, tend to be overoptimistic regarding expected returns, and are downward biased when forecasting the tails of the return distribution. A natural question that arises is whether it is easy to "turn off" these biases. We contend that completely removing behavioral biases from LLMs will be difficult. The issue is not that LLMs are unaware of investor biases. For example, when asked "What behavioral biases do investors make when using historical returns to predict future returns," ChatGPT-4's response includes Extrapolation Bias, Recency Bias, Overconfidence, Confirmation Bias, Hindsight Bias, and the Availability Heuristic, and it offers definitions of each. ChatGPT-4 can also easily summarize the evidence on short-term return reversals.

The challenge is that these biases are deeply rooted in the training process. LLMs are trained on vast datasets that reflect the full spectrum of human thought, including the biases and heuristics that are prevalent in financial discussions. The volume of data required to train an LLM makes it difficult to eliminate all instances of bias without impairing the model's ability to generate coherent and contextually appropriate responses. Consequently, even if an LLM "understands" what these biases are in theory, its outputs may still reflect these biases because they are embedded in the data from which the model learns.[13]

---

[13] For example, adding the preamble "You are a sophisticated Hedge Fund investor" to the GPT4 prompt has negligible effect on the forecast characteristics.

On the other hand, LLMs that are pre-trained on a broad corpus can be fine-tuned by retraining on a more specific datasets tailored to a particular setting, which can improve performance for domain tasks. Moreover, LLMs can also be encouraged to generate code when presented with data, further improving their data analysis capabilities. However, when an LLM faces a task without well-answered examples in the fine-tuning data, general knowledge that is susceptible to biases will likely play a larger role.

Ultimately, the findings suggest that while LLMs may offer enhanced numeracy and risk assessment capabilities compared to human counterparts, they remain vulnerable to the same cognitive biases that affect human decision-making. This underscores the importance of critically evaluating LLM-generated forecasts and considering the broader implications of integrating AI into financial decision-making processes. As these models continue to evolve, ongoing research will be crucial in developing more sophisticated methods for bias detection and mitigation, ensuring that LLMs can serve as reliable tools in finance without perpetuating harmful biases.

## 4. Conclusion

AI's capacity for objectively analyzing vast amounts of information has the potential to revolutionize financial decision-making. However, large language models (LLMs) and other AI algorithms are trained on human output, which raises the risk of embedding detrimental cognitive biases that are present in human decision-making. This study examines whether OpenAI's GPT-4 manifests behavioral biases when provided with historical return data and prompted for stock return forecasts. In particular, we assess how ChatGPT reacts to the timing and relative magnitude of individual historical returns when forming return forecasts.

Our empirical analysis indicates that ChatGPT and human forecasts rely on past data in surprisingly similar ways, with a positive, gradually declining emphasis on lagged returns. This pattern is not present in realized returns, which instead tend to exhibit a pattern of short-term reversals. The behavior of LLM forecasts is consistent with documented excessive extrapolative expectations in human decision-making. The evidence holds when using an alternative LLM (Claude) and also when providing price charts instead of numerical return data.

Additionally, ChatGPT tends to predict higher returns than both historical means and realized outcomes, indicating that its training data may have embedded an overly positive outlook on stock performance. In contrast, LLMs appear more pessimistic when forecasting the tails of the return distribution. We observe that while forecasts of $10^{th}$ percentile outcomes were more negative when compared with historical data, $90^{th}$ percentile forecasts were also lower than the upper percentiles of the historical distribution, suggesting a skewed interpretation of potential risks and rewards.

The findings contribute to the broader discourse on AI integration in financial decision-making, highlighting the need to address potential biases in LLM-generated forecasts. While LLMs show promise in assessing risk, the study cautions against assuming that these models interpret numeric data with fully rational statistical rigor. Our analysis adds to the growing literature on how LLMs replicate human behavior in financial contexts and underscores the importance of critically evaluating AI's role in finance.

# Appendix A

*A.1 Forecast Variables*

- Forecerank$_{i,t}$ – The end-of-week-*t* consensus ranking based on investors' average expectation regarding the performance of stock *i* over week *t* + 1. The rank ranges from 1 to 10 based on the ten stocks in each Forcerank contest. Source: Estimize.
- ChatGPT Rank$_{i,t}$ – ChatGPT-4's ranking of the performance of stock *i* for week *t* when provided with historical return data. Source: ChatGPT-4 prompts.
  - ChatGPT Rank$_{i,t}^{12w}$ – 12 weeks of historical return data are included in the prompt.
  - ChatGPT Rank$_{i,t}^{24w}$ – 24 weeks of historical return data are included in the prompt.
  - ChatGPT Rank$_{i,t}^{12w\ Chart}$ – An image of a price chart with 12 weeks of historical return data is included in the prompt.
  - ChatGPT Rank$_{i,t}^{24w\ Chart}$ – An image of a price chart with 24 weeks of historical return data is included in the prompt.
- Predicted ChatGPT Rank$_{i,t}$ – The fitted value obtained from regressing ChatGPT rank on lagged returns as in Specification (2) of Table 5
- Residual ChatGPT Rank$_{i,t}$ – The residual value obtained from regressing ChatGPT rank on lagged returns as in Specification (2) of Table 5.
- *AAII Sentiment$_t$* – The American Association of Individual Investors Bull – Bear Spread, defined as the fraction of survey respondents at the end of the last week of month *t* that feel the direction of the stock market over the next six months will be up (bullish), less the fraction of survey respondents that feel the direction of the stock market over the next six months will be down (bearish). Note "no change" (neutral) is also a survey option. Source: Bloomberg.
- *ChatGPT Sentiment$_t$* – ChatGPT-4's market sentiment score for month *t* when provided with 12 lagged monthly returns for the S&P 500 index and asked, "Do you feel the direction of the stock market over the next six months will be up (bullish), no change (neutral) or down (bearish)?" A score of 1 represents bullish sentiment, 0 represents neutral sentiment, and -1 represents bearish sentiment. Source: ChatGPT-4 prompts.
- ChatGPT Low$_{i,t}$ – ChatGPT-4's response to the prompt "There is a 1-in-10 chance the actual return will be less than x%," for stock *i* in month *t*+1, when provided with up to ten (but no fewer than five) years of monthly stock returns. Source: ChatGPT-4 prompts.
- ChatGPT Expected$_{i,t}$ – ChatGPT-4's response to the prompt "I expected the next month's return to be x%" for stock *i* in month *t*+1, when provided with up to ten (but no fewer than five) years. Source: ChatGPT-4 prompts.
- ChatGPT High$_{i,t}$ – ChatGPT-4's response to the prompt "There is a 1-in-10 chance the actual return will be greater than x%," for stock *i* in month *t*+1, when provided with up to ten (but and no fewer than five) years of monthly stock returns. Source: ChatGPT-4 prompts.

*A.2 Return Measures*

- Return$_{i,t}$ – Return for stock *i* in week or month *t*. Source: CRSP
- Return Rank$_{i,t}$ – Stock return performance rank for stock *t* in week *t*. The rank is from 1 to 10 for the ten stocks in the Forcerank contest.
- S&P Return$_t$ – Return for the S&P 500 Index for month *t*. Source: CRSP.

*A.3 Control Variables*

- *Market Capitalization*$_{i,t}$ – The market value of equity measured for month *t*. Source: CRSP.
- *Book to Market*$_{i,t}$ – The sum of the market value of equity and total debts, expressed as a fraction of total assets, measured for the fiscal year prior to month *t*. Source: Compustat.
- *Asset Growth*$_{i,t}$ – The percentage change in book value of total assets from balance sheet, measured for the fiscal year prior to month *t*. Source: Compustat.
- *Profitability*$_{i,t}$ – Revenue minus cost of goods sold, divided by total assets. Measured for the fiscal year prior to month *t*. Source: Compustat.
- *Market Beta*$_{i,t}$ – Market beta from fitting the CAPM to daily stock returns for stock *i* in month *t*. Source: CRSP.
- *Return MAX*$_{i,t}$ – The maximum daily return for stock *i* in month *t*. Source: CRSP.
- *Turnover*$_{i,t}$ – The sum of daily dollar volume over market cap for stock *i* in week *t*. Source: CRSP.

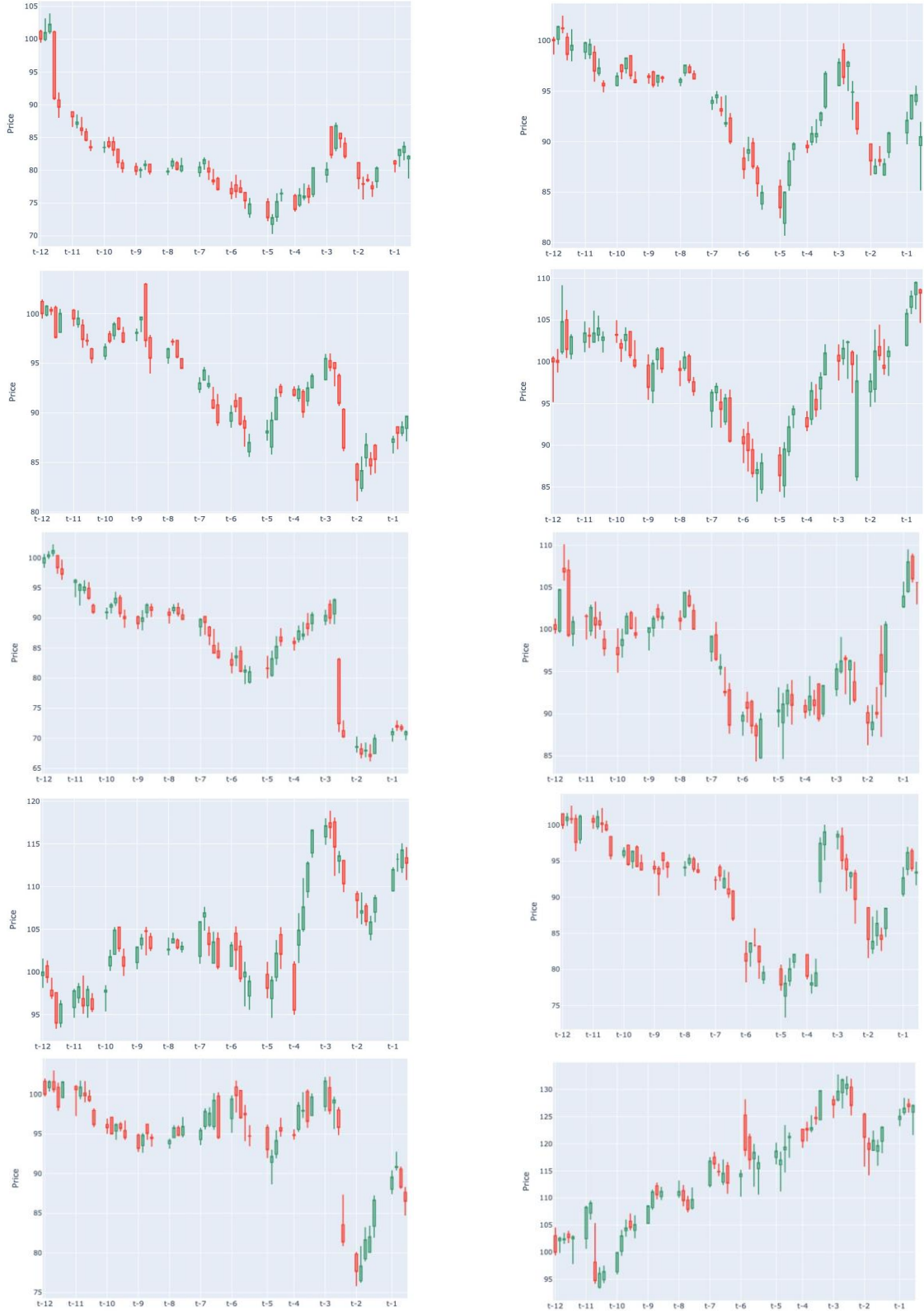

**Figure 1. Price Charts for Forcerank Contest Stocks.** The plots show an example set of 12-week price charts for Forcerank contest stocks. For each Forcerank contest, we provide a corresponding set of historical price figures to ChatGPT-4 and prompt it to issue performance rankings for the ten contest stocks over the following week.

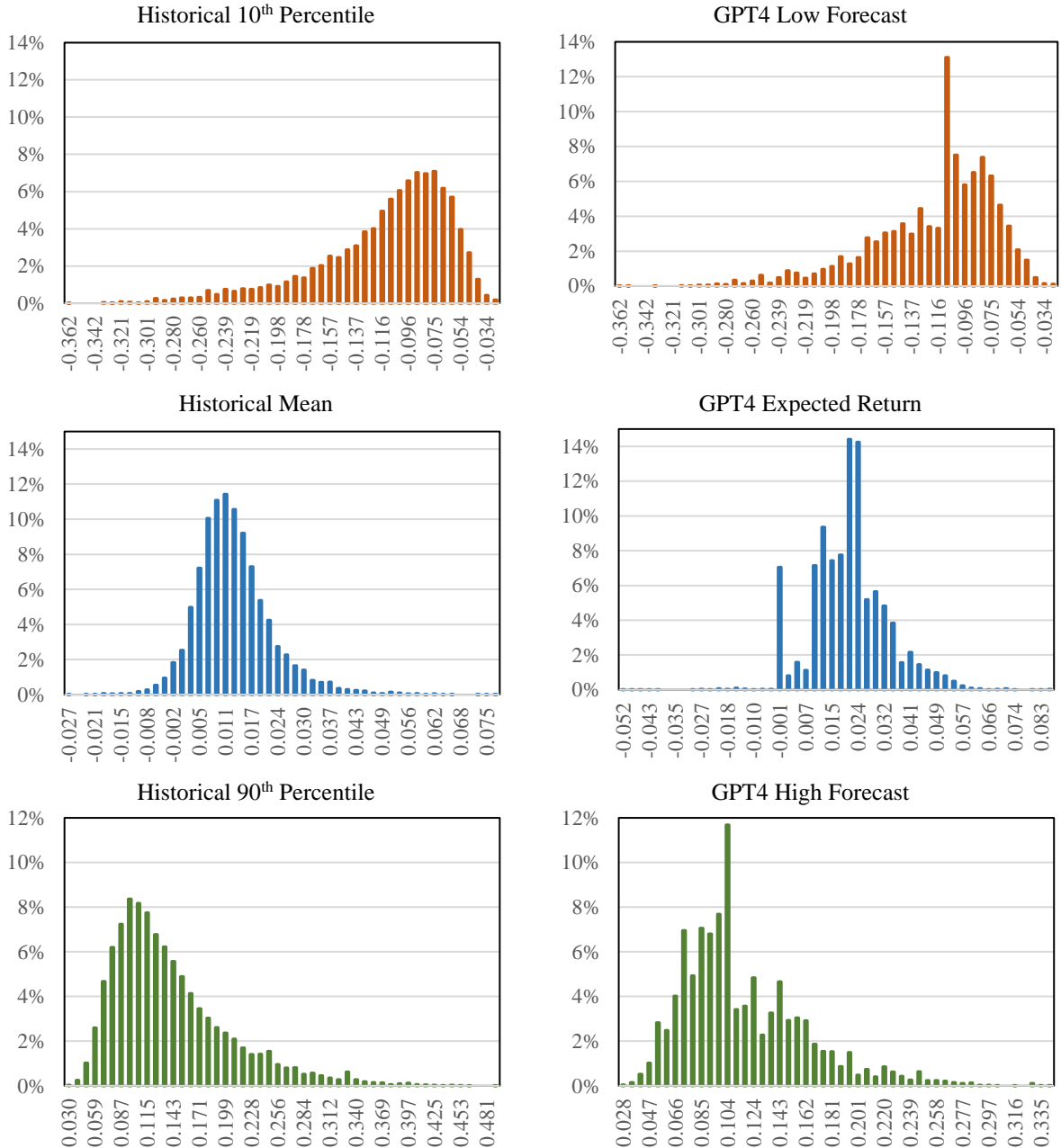

**Figure 2. Historical and LLM Forecasts of Low, Expected, and High returns.** ChatGPT-4 is provided with up to ten (no fewer than five) years of historical monthly returns for a randomly chosen stock-month, and the process is repeated 10,000 times. The left plots show the distribution of the 10th percentile, mean, and 90th percentile of the historical samples provided to ChatGPT-4. The right plots show the distribution of the resulting next-month 10th percentile, expected, and 90th percentile return forecasts produced by ChatGPT.

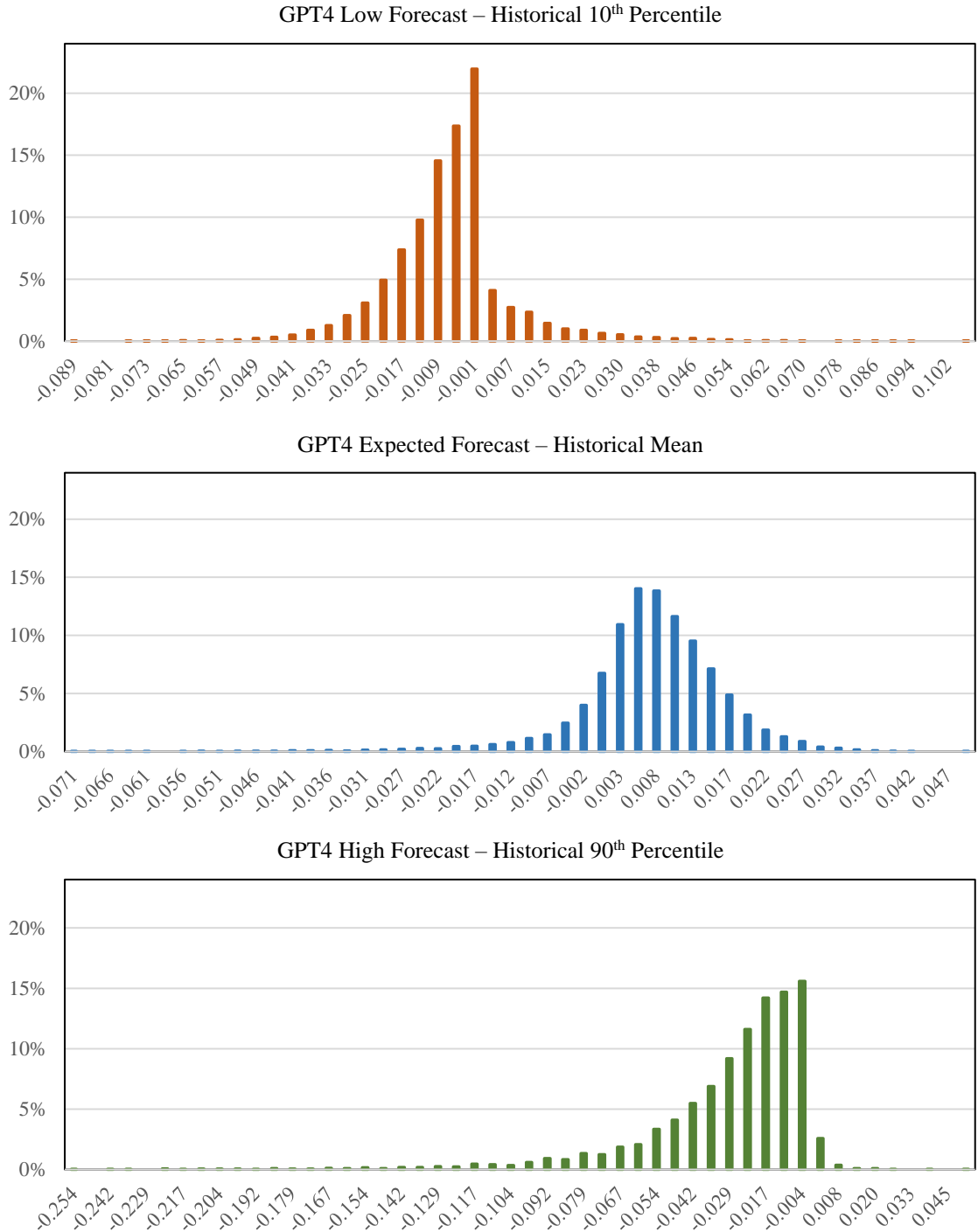

**Figure 3. Differences between Historical and ChatGPT Forecasts.** ChatGPT-4 is provided with up to ten (no fewer than five) years of historical monthly returns for a randomly chosen stock-month, and the process is repeated 10,000 times. The top (lower) plot shows the distribution of the differences between the 10$^{th}$ (90$^{th}$) percentile of the historical sample provided to ChatGPT-4 and the resulting next-month 10$^{th}$ (90$^{th}$) percentile forecast produced by ChatGPT. The middle panel plots the distribution of the differences between the historical mean and ChatGPT's expected return forecast.

**Table 1. Sample Statistics**

The table presents sample descriptive statistics. Panel A provides contest-stock level statistics for the sample of 1,286 Forcerank contests that occurred between February 2016 and December 2017. Panel B provides statistics for monthly observations from the American Association of Individual Investor survey sample, covering July 1987 through June 2024. In each panel we also include statistics for ChatGPT-4 produced performance rank and sentiment forecasts. Detailed definitions can be found in the Appendix.

Panel A: Contest-Stock level sample

|  | Obs. | Mean | Standard Deviation | 25th Percentile | Median | 75th Percentile |
|---|---|---|---|---|---|---|
| Realized Returns (%) | 12,719 | 0.43 | 4.25 | -1.61 | 0.30 | 2.38 |
| Forcerank Score | 12,719 | 5.54 | 2.86 | 3.00 | 6.00 | 8.00 |
| ChatGPT-12week | 12,719 | 5.54 | 2.85 | 3.00 | 6.00 | 8.00 |
| ChatGPT-24week | 12,719 | 5.54 | 2.85 | 3.00 | 6.00 | 8.00 |
| Market Capitalization | 12,719 | 9.75 | 1.66 | 8.36 | 9.76 | 10.98 |
| Book to Market | 10,872 | -1.46 | 0.93 | -1.95 | -1.40 | -0.85 |
| Asset Growth | 10,872 | 0.12 | 0.35 | -0.02 | 0.06 | 0.17 |
| Profitability | 10,872 | 0.35 | 0.23 | 0.20 | 0.33 | 0.46 |
| Market Beta | 10,872 | 1.26 | 0.93 | 0.71 | 1.15 | 1.68 |
| Return MAX | 10,872 | 0.04 | 0.03 | 0.02 | 0.03 | 0.04 |
| Turnover | 10,872 | 0.48 | 2.44 | 0.03 | 0.05 | 0.08 |

Panel B: Monthly sample

|  | Obs. | Mean | Standard Deviation | 25th Percentile | Median | 75th Percentile |
|---|---|---|---|---|---|---|
| S&P 500 Returns (%) | 438 | 0.73 | 4.40 | -1.78 | 1.17 | 3.52 |
| AAII Bull – Bear (%) | 438 | 0.06 | 0.18 | -0.07 | 0.06 | 0.20 |
| GPT4 Sentiment (%) | 438 | 0.37 | 0.72 | 0.00 | 0.00 | 1.00 |

**Table 2. Extrapolation of Past Returns: Humans, ChatGPT, and Realized Returns**

This table presents the results from linear regressions at the contest-stock-week level, as specified in Eq. (1) in the main text. Specification (1) uses the consensus Forcerank ranking (ranging from one to ten) as the dependent variable, representing the average ranking of a stock across all participants in a contest, with ten indicating the highest rank and one the lowest. In specifications (2) to (4), the dependent variable is the stock ranking generated by ChatGPT-4 based on stock returns in the past 12 weeks. Specification (5) focuses on one-week-ahead stock returns as the dependent variable. The explanatory variables include lagged returns from week $t-11$ to week t. Standard errors are clustered by contest, and *, **, and *** denote statistical significance at the 10%, 5%, and 1% levels, respectively. The sample period is from February 2016 to December 2017.

|  | Forecerank$_{i,t}$ | | ChatGPT Rank$_{i,t}^{12w}$ | | Return$_{i,t+1}$ |
|---|---|---|---|---|---|
|  | (1) | (2) | (3) | (4) | (5) |
| Forecerank$_{i,t}$ |  |  | 0.28*** | 0.17*** |  |
|  |  |  | (27.53) | (21.33) |  |
| Return$_{i,t}$ | 12.62*** | 38.84*** |  | 36.66*** | 1.15 |
|  | (20.12) | (38.73) |  | (37.64) | (0.79) |
| Return$_{i,t-1}$ | 2.31*** | 2.94*** |  | 2.54*** | -3.06** |
|  | (4.36) | (4.91) |  | (4.43) | (-2.31) |
| Return$_{i,t-2}$ | 2.18*** | 1.92*** |  | 1.54*** | -3.90*** |
|  | (4.22) | (3.38) |  | (2.83) | (-2.78) |
| Return$_{i,t-3}$ | 2.58*** | 0.87 |  | 0.43 | -2.80* |
|  | (5.00) | (1.57) |  | (0.80) | (-1.93) |
| Return$_{i,t-4}$ | 2.25*** | 1.15* |  | 0.76 | 0.84 |
|  | (4.26) | (1.95) |  | (1.33) | (0.61) |
| Return$_{i,t-5}$ | 2.32*** | 2.08*** |  | 1.68*** | 3.07** |
|  | (4.46) | (3.72) |  | (3.14) | (2.26) |
| Return$_{i,t-6}$ | 1.53*** | -0.38 |  | -0.64 | 5.32*** |
|  | (3.03) | (-0.65) |  | (-1.16) | (3.54) |
| Return$_{i,t-7}$ | 1.32** | 0.42 |  | 0.19 | -1.35 |
|  | (2.53) | (0.68) |  | (0.33) | (-0.97) |
| Return$_{i,t-8}$ | 1.22** | 0.72 |  | 0.51 | -1.59 |
|  | (2.42) | (1.24) |  | (0.93) | (-1.08) |
| Return$_{i,t-9}$ | 0.83* | 0.64 |  | 0.50 | -3.04** |
|  | (1.68) | (1.14) |  | (0.92) | (-2.16) |
| Return$_{i,t-10}$ | 1.57*** | 1.87*** |  | 1.60*** | -0.93 |
|  | (3.29) | (3.36) |  | (2.99) | (-0.64) |
| Return$_{i,t-11}$ | 0.23 | 1.10** |  | 1.06** | -1.79 |
|  | (0.49) | (2.09) |  | (2.09) | (-1.37) |
| Observations | 12,668 | 12,668 | 12,719 | 12,668 | 12,668 |
| R-squared | 0.043 | 0.353 | 0.079 | 0.381 | 0.011 |

**Table 3. Asymmetric Extrapolation of Positive and Negative Returns.**
This table presents the results from linear regressions in which the explanatory variables are the positive and negative components of 12 weekly return lags. The positive component of a stock return is defined as Max(Return,0) and the negative component is defined as Min(Return,0). Specification (1) uses the consensus Forcerank ranking (ranging from one to ten) as the dependent variable, representing the average ranking of a stock across all participants in a contest, with ten indicating the highest rank and one the lowest. In Specification (2), the dependent variable is the stock ranking generated by ChatGPT-4 based on past 12 weekly returns, and in Specification (3), the stock ranking generated by ChatGPT-4 is based on 24 weekly returns. Standard errors are clustered by contest, and *, **, and *** denote statistical significance at the 10%, 5%, and 1% levels, respectively. The sample period is from February 2016 to December 2017.

|  | (1) Forcerank$_{i,t}$ | | (2) ChatGPT Rank$_{i,t}^{12w}$ | | (3) ChatGPT Rank$_{i,t}^{24w}$ | |
|---|---|---|---|---|---|---|
|  | Positive Returns | Negative Returns | Positive Returns | Negative Returns | Positive Returns | Negative Returns |
| Return$_{i,t}$ | 9.50*** | 16.71*** | 41.11*** | 36.07*** | 39.23*** | 26.30*** |
|  | (9.77) | (13.31) | (23.66) | (21.09) | (24.16) | (17.68) |
| Return$_{i,t-1}$ | -0.01 | 5.12*** | 4.19*** | 1.26 | 15.80*** | 1.68* |
|  | (-0.02) | (4.78) | (4.36) | (1.31) | (13.41) | (1.81) |
| Return$_{i,t-2}$ | -1.03 | 6.28*** | 0.95 | 3.17*** | 2.38*** | 2.47** |
|  | (-1.20) | (6.71) | (1.17) | (3.00) | (2.90) | (2.32) |
| Return$_{i,t-3}$ | 0.57 | 5.21*** | 0.32 | 1.64* | 1.64** | 2.69*** |
|  | (0.64) | (5.33) | (0.39) | (1.69) | (2.02) | (2.75) |
| Return$_{i,t-4}$ | -0.06 | 5.81*** | 0.44 | 2.17** | 5.41*** | 2.87*** |
|  | (-0.07) | (5.15) | (0.50) | (2.07) | (5.22) | (2.76) |
| Return$_{i,t-5}$ | -1.74* | 6.72*** | 1.00 | 3.55*** | 5.09*** | 4.04*** |
|  | (-1.94) | (6.90) | (1.20) | (3.52) | (5.35) | (4.11) |
| Return$_{i,t-6}$ | -0.96 | 4.13*** | -0.05 | -0.49 | 2.21*** | 0.02 |
|  | (-1.15) | (4.02) | (-0.06) | (-0.45) | (2.73) | (0.01) |
| Return$_{i,t-7}$ | -1.65** | 4.89*** | 1.64* | -1.00 | 1.86** | 0.92 |
|  | (-2.00) | (4.69) | (1.90) | (-0.87) | (2.22) | (0.80) |
| Return$_{i,t-8}$ | -1.49* | 4.40*** | -0.72 | 2.52** | -1.01 | 3.90*** |
|  | (-1.75) | (4.53) | (-0.82) | (2.32) | (-1.18) | (3.83) |
| Return$_{i,t-9}$ | -1.40* | 3.14*** | -0.37 | 1.89** | -0.31 | 1.79* |
|  | (-1.71) | (3.09) | (-0.42) | (1.97) | (-0.37) | (1.95) |
| Return$_{i,t-10}$ | -0.70 | 3.73*** | 1.11 | 2.67*** | 0.59 | 3.10*** |
|  | (-0.85) | (3.49) | (1.26) | (2.86) | (0.69) | (2.98) |
| Return$_{i,t-11}$ | -1.57* | 1.89* | 1.34 | 0.66 | 0.56 | 1.30 |
|  | (-1.93) | (1.91) | (1.61) | (0.78) | (0.67) | (1.50) |
| Observations | 12,719 | | 12,719 | | 12,719 | |
| R-squared | 0.073 | | 0.356 | | 0.305 | |

**Table 4. Extrapolation of Past Return Ranks**

This table repeats the regression analysis in Table 2, employing return ranks as the explanatory variables (i.e., the stocks' actual past rankings converted from past weekly returns in the contest). Specification (1) uses the consensus Forcerank ranking (ranging from one to ten) as the dependent variable, representing the average ranking of a stock across all participants in a contest, with ten indicating the highest rank and one the lowest. In specifications (2) to (4), the dependent variable is the stock ranking generated by ChatGPT-4 based on unadjusted stock returns in the past 12 weeks. Specification (5) focuses on the realized one-week-ahead stock ranks as the dependent variable. The explanatory variables include lagged returns from week $t-11$ to week $t$. Standard errors are clustered by contest, and *, **, and *** denote statistical significance at the 10%, 5%, and 1% levels, respectively. The sample period is from February 2016 to December 2017.

|  | Forcerank$_{i,t}$ | ChatGPT Rank$_{i,t}^{12w}$ | | | Ret Rank$_{i,t+1}$ |
|---|---|---|---|---|---|
|  | (1) | (2) | (3) | (4) | (5) |
| Forcerank$_{i,t}$ |  |  | 0.28*** | 0.00 |  |
|  |  |  | (27.53) | (1.01) |  |
| Return Rank$_{i,t}$ | 0.30*** | 0.91*** |  | 0.91*** | -0.02 |
|  | (29.26) | (222.18) |  | (215.74) | (-1.49) |
| Return Rank$_{i,t-1}$ | 0.04*** | 0.07*** |  | 0.07*** | -0.01 |
|  | (4.81) | (16.30) |  | (16.26) | (-1.28) |
| Return Rank$_{i,t-2}$ | 0.05*** | 0.02*** |  | 0.02*** | -0.02* |
|  | (5.06) | (4.40) |  | (4.33) | (-1.79) |
| Return Rank$_{i,t-3}$ | 0.05*** | 0.01*** |  | 0.01** | -0.00 |
|  | (5.34) | (2.61) |  | (2.55) | (-0.24) |
| Return Rank$_{i,t-4}$ | 0.04*** | 0.01*** |  | 0.01*** | 0.02** |
|  | (4.68) | (3.56) |  | (3.51) | (2.43) |
| Return Rank$_{i,t-5}$ | 0.04*** | 0.01*** |  | 0.01*** | 0.02 |
|  | (3.86) | (3.37) |  | (3.33) | (1.50) |
| Return Rank$_{i,t-6}$ | 0.03*** | 0.01** |  | 0.01** | 0.03*** |
|  | (3.44) | (2.38) |  | (2.34) | (3.08) |
| Return Rank$_{i,t-7}$ | 0.03*** | 0.01*** |  | 0.01*** | 0.00 |
|  | (3.37) | (2.84) |  | (2.80) | (0.24) |
| Return Rank$_{i,t-8}$ | 0.02*** | 0.01 |  | 0.01 | 0.00 |
|  | (2.62) | (1.52) |  | (1.49) | (0.17) |
| Return Rank$_{i,t-9}$ | 0.02** | 0.01** |  | 0.01** | 0.02* |
|  | (2.31) | (2.18) |  | (2.16) | (1.72) |
| Return Rank$_{i,t-10}$ | 0.02** | 0.01** |  | 0.01** | -0.01 |
|  | (2.41) | (2.37) |  | (2.34) | (-0.96) |
| Return Rank$_{i,t-11}$ | 0.01 | 0.02*** |  | 0.02*** | -0.01 |
|  | (1.22) | (4.70) |  | (4.69) | (-0.86) |
| Observations | 12,719 | 12,719 | 12,719 | 12,719 | 12,719 |
| R-squared | 0.102 | 0.836 | 0.079 | 0.836 | 0.003 |

**Table 5. LLM Extrapolation of Price Charts**

This table presents the results from linear regressions at the contest-stock-week level, as specified in Eq. (1) in the main text. Forcerank is the average ranking of a stock across all participants in a contest, with ten indicating the highest rank and one the lowest. The dependent variable is the performance ranking generated by ChatGPT-4 based on price charts over the past 12 weeks. The explanatory variables include lagged returns relative to forecast week $t+1$. Standard errors are clustered by contest, and *, **, and *** denote statistical significance at the 10%, 5%, and 1% levels, respectively. The sample period is from February 2016 to December 2017.

|  | ChatGPT Rank$_{i,t}^{12w\ Chart}$ | | |
| --- | --- | --- | --- |
|  | (1) | (2) | (3) |
| Forcerank$_{i,t}$ |  | 0.05*** | 0.02** |
|  |  | (5.06) | (2.43) |
| Return$_{i,t}$ | 6.29*** |  | 6.00*** |
|  | (11.62) |  | (11.11) |
| Return$_{i,t-1}$ | 5.00*** |  | 4.94*** |
|  | (9.67) |  | (9.57) |
| Return$_{i,t-2}$ | 3.73*** |  | 3.68*** |
|  | (6.93) |  | (6.84) |
| Return$_{i,t-3}$ | 2.79*** |  | 2.73*** |
|  | (5.21) |  | (5.11) |
| Return$_{i,t-4}$ | 2.33*** |  | 2.28*** |
|  | (4.60) |  | (4.51) |
| Return$_{i,t-5}$ | 2.07*** |  | 2.01*** |
|  | (3.93) |  | (3.84) |
| Return$_{i,t-6}$ | 2.45*** |  | 2.41*** |
|  | (4.71) |  | (4.64) |
| Return$_{i,t-7}$ | 1.81*** |  | 1.78*** |
|  | (3.47) |  | (3.42) |
| Return$_{i,t-8}$ | 1.23** |  | 1.20** |
|  | (2.53) |  | (2.48) |
| Return$_{i,t-9}$ | 0.63 |  | 0.61 |
|  | (1.30) |  | (1.26) |
| Return$_{i,t-10}$ | 1.81*** |  | 1.77*** |
|  | (3.57) |  | (3.51) |
| Return$_{i,t-11}$ | 0.79* |  | 0.78* |
|  | (1.69) |  | (1.68) |
| Observations | 12,668 | 12,719 | 12,668 |
| R-squared | 0.024 | 0.002 | 0.025 |

**Table 6. Extrapolative beliefs: Exponential decay model.**

The table presents the results of a contest-level nonlinear regression specified in Eq. (3) of the main text:

$$Y_{i,t} = 5.5 + \lambda_1 \cdot \sum_{s=0}^{12} w_s R_{i,t-s} + \epsilon_{i,t}, \quad \text{where } w_s = \frac{\lambda_2^s}{\sum_{j=0}^{12} \lambda_2^j}$$

In Specification (1) the dependent variable is the consensus ranking (one to ten) representing a stock's average ranking across all contest participants. In Specification (2), the dependent variable is the ranking produced by ChatGPT4 using 12 weekly return lags, and in Specification (3), 24 return lags are considered. The explanatory variables include lagged returns from week $t-11$ to week $t$. The exponential decay model is estimated using GMM, and *, **, and *** denote statistical significance at the 10%, 5%, and 1% levels, respectively. DHJ show theoretically that a higher $l_1$ and a lower $l_2$ jointly lead to a higher degree of extrapolation and $l_1(1-l_2)$ represents the degree of extrapolation. The sample period is from February 2016 to December 2017.

|  | Forecerank$_{i,t}$ | ChatGPT Rank$_{i,t}^{12w}$ | ChatGPT Rank$_{i,t}^{24w}$ |
|---|---|---|---|
|  | (1) | (2) | (3) |
| $l_1$ | 16.98*** | 40.72*** | 45.68*** |
|  | (15.53) | (48.89) | (45.30) |
| $l_2$ | 0.28*** | 0.07*** | 0.27*** |
|  | (6.78) | (5.16) | (18.55) |
| $l_1(1-l_2)$ | 12.19 | 38.03 | 33.21 |

**Table 7. ChatGPT Forecast Ranks and Future Stock Returns**

This table presents the results from Fama-MacBeth return forecasting regressions. For each week $t$ and each stock $i$, the dependent variable is the daily return of stock $i$ over week $t+1$. The return predictors include the ChatGPT-4 stock rank and its decomposed components: the predicted component is derived as the fitted value from the nonlinear regression specified in Equation (3), while the residual component is referred to as the residual ChatGPT rank. Panel A focuses on the ranking produced by ChatGPT4 using 12 weekly return lags, and Panel B focuses on ChatGPT4 ranks based on 24 return lags. Control variables, measured at week $t$, include log market capitalization, log book-to-market, asset growth, gross profits-to-assets, market beta, weekly turnover, and the max daily return in the last month. Returns are measured in basis points, with t-statistics provided in parentheses. *, **, and *** denote significance at the 10%, 5%, and 1% levels, respectively. The sample period is from February 2016 to December 2017.

Panel A: Forecast inferred from 12 lagged returns

|  | (1) | (2) | (3) | (4) | (5) | (6) |
|---|---|---|---|---|---|---|
| ChatGPT Rank$_{i,t}^{12w}$ | -0.12 |  |  | -0.41 |  |  |
|  | (-0.35) |  |  | (-1.15) |  |  |
| Predicted ChatGPT Rank$_{i,t}^{12w}$ |  | -0.38 |  |  | -1.08* |  |
|  |  | (-0.76) |  |  | (-1.92) |  |
| Residual ChatGPT Rank$_{i,t}^{12w}$ |  |  | 0.11 |  |  | 0.35 |
|  |  |  | (0.18) |  |  | (0.56) |
| Controls | No | No | No | Yes | Yes | Yes |
| Observations | 58,056 | 58,056 | 58,056 | 49,683 | 49,683 | 49,683 |
| R-squared | 0.012 | 0.017 | 0.013 | 0.183 | 0.189 | 0.184 |

Panel B: Forecast inferred from 24 lagged returns

|  | (1) | (2) | (3) | (4) | (5) | (6) |
|---|---|---|---|---|---|---|
| ChatGPT Rank$_{i,t}^{24w}$ | -0.50 |  |  | -0.78** |  |  |
|  | (-1.40) |  |  | (-2.05) |  |  |
| Predicted ChatGPT Rank$_{i,t}^{24w}$ |  | -0.69 |  |  | -1.34** |  |
|  |  | (-1.25) |  |  | (-2.20) |  |
| Residual ChatGPT Rank$_{i,t}^{24w}$ |  |  | -0.67 |  |  | -0.67 |
|  |  |  | (-1.35) |  |  | (-1.18) |
| Controls | No | No | No | Yes | Yes | Yes |
| Observations | 58,056 | 58,056 | 58,056 | 49,683 | 49,683 | 49,683 |
| R-squared | 0.013 | 0.017 | 0.012 | 0.184 | 0.188 | 0.185 |

**Table 8. Market Return Extrapolation: AAII and ChatGPT Sentiment**

This table presents the results from linear regressions at the month level. In Specification (1), the dependent variable is AAII sentiment, measured as the percentage of "bearish" investors minus the percentage of "bullish" investors in the last week of each month. In Specifications (2) to (4), the dependent variable is the ChatGPT-4 sentiment generated based on the US stock market (S&P 500) returns in the past 12 months. *, **, and *** denote statistical significance at the 10%, 5%, and 1% levels, respectively. The sample period is from July 1987 to June 2024.

|  | AAII Sentiment$_t$ |  | ChatGPT Sentiment$_t$ |  |
| --- | --- | --- | --- | --- |
|  | (1) | (2) | (3) | (4) |
| AAII Sent$_t$ |  |  | 1.63*** | 0.30** |
|  |  |  | (8.92) | (2.57) |
| S&P Return$_t$ | 1.42*** | 9.17*** |  | 8.74*** |
|  | (6.60) | (16.68) |  | (15.14) |
| S&P Return$_{t-1}$ | 0.80*** | 5.38*** |  | 5.14*** |
|  | (3.99) | (10.20) |  | (9.38) |
| S&P Return$_{t-2}$ | 0.38** | 3.99*** |  | 3.87*** |
|  | (1.99) | (8.78) |  | (8.41) |
| S&P Return$_{t-3}$ | 0.35* | 3.65*** |  | 3.55*** |
|  | (1.77) | (7.48) |  | (7.26) |
| S&P Return$_{t-4}$ | 0.26 | 2.68*** |  | 2.60*** |
|  | (1.35) | (4.89) |  | (4.76) |
| S&P Return$_{t-5}$ | 0.22 | 3.39*** |  | 3.32*** |
|  | (1.14) | (7.04) |  | (6.79) |
| S&P Return$_{t-6}$ | 0.12 | 2.48*** |  | 2.44*** |
|  | (0.60) | (6.03) |  | (5.88) |
| S&P Return$_{t-7}$ | 0.18 | 2.00*** |  | 1.94*** |
|  | (0.93) | (4.43) |  | (4.35) |
| S&P Return$_{t-8}$ | 0.33 | 1.78*** |  | 1.68*** |
|  | (1.63) | (3.69) |  | (3.53) |
| S&P Return$_{t-9}$ | 0.10 | 2.13*** |  | 2.10*** |
|  | (0.50) | (4.83) |  | (4.81) |
| S&P Return$_{t-10}$ | 0.25 | 1.47*** |  | 1.39*** |
|  | (1.40) | (3.26) |  | (3.08) |
| S&P Return$_{t-11}$ | 0.04 | 2.12*** |  | 2.11*** |
|  | (0.21) | (5.11) |  | (5.04) |
| Observations | 438 | 438 | 438 | 438 |
| R-squared | 0.194 | 0.687 | 0.168 | 0.692 |

**Table 9. Market Return Extrapolation: Claude Sentiment**

This table presents the results from linear regressions at the month level. In Specification (1), the dependent variable is Claude sentiment generated based on the US stock market (S&P 500) returns in the past 12 months. The independent variables in AAII sentiment, measured as the percentage of "bearish" investors minus the percentage of "bullish" investors in the last week of each month, and ChatGPT sentiment. . In Specifications (2) to (4), *, **, and *** denote statistical significance at the 10%, 5%, and 1% levels, respectively. The sample period is from July 1987 to June 2024.

|  | Claude Sentiment$_t$ | | | |
|---|---|---|---|---|
|  | (1) | (2) | (3) | (4) |
| AAII Sent$_t$ |  | 1.51*** | 0.09 |  |
|  |  | (7.53) | (0.68) |  |
| ChatGPT Sent$_t$ |  |  |  | 0.81*** |
|  |  |  |  | (23.90) |
| S&P Return$_t$ | 10.64*** |  | 10.51*** |  |
|  | (17.74) |  | (16.44) |  |
| S&P Return$_{t-1}$ | 5.74*** |  | 5.67*** |  |
|  | (9.83) |  | (9.36) |  |
| S&P Return$_{t-2}$ | 3.60*** |  | 3.56*** |  |
|  | (6.52) |  | (6.43) |  |
| S&P Return$_{t-3}$ | 2.61*** |  | 2.58*** |  |
|  | (4.85) |  | (4.75) |  |
| S&P Return$_{t-4}$ | 1.92*** |  | 1.90*** |  |
|  | (3.34) |  | (3.27) |  |
| S&P Return$_{t-5}$ | 2.31*** |  | 2.29*** |  |
|  | (3.95) |  | (3.89) |  |
| S&P Return$_{t-6}$ | 1.29** |  | 1.28** |  |
|  | (2.30) |  | (2.27) |  |
| S&P Return$_{t-7}$ | 1.43*** |  | 1.41*** |  |
|  | (2.87) |  | (2.83) |  |
| S&P Return$_{t-8}$ | 1.43*** |  | 1.40** |  |
|  | (2.63) |  | (2.55) |  |
| S&P Return$_{t-9}$ | 1.05** |  | 1.04** |  |
|  | (2.00) |  | (1.99) |  |
| S&P Return$_{t-10}$ | 1.17** |  | 1.15** |  |
|  | (2.49) |  | (2.42) |  |
| S&P Return$_{t-11}$ | 0.66 |  | 0.65 |  |
|  | (1.37) |  | (1.36) |  |
| Observations | 438 | 438 | 438 | 438 |
| R-squared | 0.630 | 0.133 | 0.631 | 0.603 |

**Table 10. LLM Forecasts of the Return Distribution – Bias Tests and Calibration Evidence**

The table presents descriptive statistics and bias tests for the return distribution forecasts generated by ChatGPT-4. Panel A presents descriptive statistics for historical returns, GPT4 forecasts, and realized returns. Panel B examines bias tests, comparing the expected forecasts against historical means and realized outcomes. Panel C provides calibration evidence, evaluating the accuracy of the Low and High forecasts relative to their historical percentiles. All statistics are based on 10,000 stock-month observations randomly selected from the 1926 to 2023 period, with up to ten years of historical monthly returns provided to ChatGPT-4.

Panel A: Descriptive statistics

| Variable | Obs. | Mean | Std. | 5% | 25% | 50% | 75% | 95% |
|---|---|---|---|---|---|---|---|---|
| Expected forecast | 10,000 | 2.23 | 1.23 | 0.00 | 1.45 | 2.27 | 2.84 | 4.37 |
| Historical mean | 10,000 | 1.40 | 0.94 | 0.14 | 0.81 | 1.29 | 1.85 | 3.07 |
| Realized returns | 9954 | 1.15 | 14.61 | -15.69 | -5.42 | 0.00 | 5.71 | 20.34 |
| Low forecast | 10,000 | -11.50 | 4.66 | -20.78 | -13.97 | -10.53 | -7.97 | -5.64 |
| Historical 10% | 10,000 | -11.02 | 5.21 | -22.02 | -13.50 | -9.76 | -7.26 | -4.83 |
| High forecast | 10,000 | 11.89 | 4.55 | 6.20 | 8.68 | 10.75 | 14.54 | 20.83 |
| Historical 90% | 10,000 | 14.48 | 6.36 | 7.02 | 9.98 | 12.95 | 17.46 | 27.27 |
| Confidence interval % | 10,000 | 23.39 | 8.87 | 12.23 | 17.04 | 21.29 | 28.46 | 41.20 |
| Historical 90% - 10% | 10,000 | 25.50 | 11.25 | 12.11 | 17.43 | 22.67 | 30.88 | 48.87 |

Panel B: Forecast bias

| Difference | Mean Difference | p-Value |
|---|---|---|
| Expected forecast = Historical mean | 0.83 | 0.000 |
| Expected forecast = Realized return | 1.08 | 0.000 |
| Low forecast = Historical 10% | -0.49 | 0.000 |
| High forecast = Historical 90% | -2.60 | 0.000 |

Panel C: Realized returns relative to historical and ChatGPT forecasts

| | |
|---|---|
| % of realized returns below low forecast | 10.33 |
| % of realized returns in confidence interval | 76.53 |
| % of realized returns above high forecast | 12.64 |
| % of realized returns below historical 10% | 11.61 |
| % of realized returns in historical interval | 79.02 |
| % of realized returns above historical 90% | 9.30 |

**Table 11. ChatGPT Return Forecasts: The Role of Historical Return Magnitude**
This table presents the results from linear regressions conducted at the stock-month level, as specified in Equation (5) of the main text. The stock-month sample is constructed by randomly selecting 100 months between 1926 and 2023 and then selecting 10 stocks from each size decile, based on Fama-French size breakpoints. In Column (1), the dependent variable is the 10th percentile return forecast generated by ChatGPT-4, based on individual stock returns over the past 120 months. Columns (2) and (3) use the mean return forecast and the 90th percentile return forecast generated by ChatGPT-4 as the dependent variables, respectively. The explanatory variables are the minimum, 10$^{th}$ to 90$^{th}$ percentiles, and maximum of the 120 realized monthly returns for each stock. All specifications include year-month fixed effects. Standard errors are clustered by year-month. *, **, and *** denote significance at the 10%, 5%, and 1% levels, respectively.

| Historical Returns | ChatGPT Low$_{i,t}$ (1) | ChatGPT Expected$_{i,t}$ (2) | ChatGPT High$_{i,t}$ (3) |
|---|---|---|---|
| Minimum Return | 0.01*** | 0.01*** | 0.00 |
|  | (6.30) | (26.23) | (1.66) |
| 10$^{th}$ Percentile | 0.81*** | 0.10*** | -0.14*** |
|  | (52.88) | (27.36) | (-8.83) |
| 20$^{th}$ Percentile | -0.05** | 0.09*** | 0.01 |
|  | (-1.99) | (17.71) | (0.28) |
| 30$^{th}$ Percentile | -0.01 | 0.08*** | -0.04 |
|  | (-0.46) | (15.01) | (-1.08) |
| 40$^{th}$ Percentile | 0.00 | 0.09*** | -0.02 |
|  | (0.08) | (11.44) | (-0.48) |
| 50$^{th}$ Percentile | 0.03 | 0.10*** | -0.03 |
|  | (0.98) | (12.24) | (-0.76) |
| 60$^{th}$ Percentile | -0.02 | 0.08*** | 0.04 |
|  | (-0.92) | (9.12) | (0.99) |
| 70$^{th}$ Percentile | -0.02 | 0.08*** | -0.05 |
|  | (-0.80) | (13.26) | (-1.25) |
| 80$^{th}$ Percentile | -0.05*** | 0.10*** | 0.49*** |
|  | (-3.44) | (21.19) | (20.54) |
| 90$^{th}$ Percentile | -0.04*** | 0.13*** | 0.32*** |
|  | (-4.66) | (51.27) | (18.49) |
| Maximum Return | 0.00*** | 0.01*** | -0.00 |
|  | (4.00) | (37.05) | (-0.13) |
| Year-Month Fixed Effects | Yes | Yes | Yes |
| Observations | 10,000 | 10,000 | 10,000 |
| R-squared | 0.937 | 0.907 | 0.864 |

**Table IA1. Extrapolation of Past Returns: Contest-Adjusted Returns**

This table repeats the regression analysis in Table 2, focusing on contest-adjusted returns (i.e., the stock return in excess of the average return of the ten stocks in the contest). Specification (1) uses the consensus Forcerank ranking (ranging from one to ten) as the dependent variable, representing the average ranking of a stock across all participants in a contest, with ten indicating the highest rank and one the lowest. In specifications (2) to (4), the dependent variable is the stock ranking generated by ChatGPT-4 based on unadjusted stock returns in the past 12 weeks. Specification (5) focuses on one-week-ahead stock contest-adjusted returns as the dependent variable. The explanatory variables include lagged returns from week $t − 11$ to week $t$. Standard errors are clustered by contest, and *, **, and *** denote statistical significance at the 10%, 5%, and 1% levels, respectively.

|  | Forcerank$_{i,t}$ | | ChatGPT Rank$_{i,t}^{12w}$ | | Adj Return$_{i,t+1}$ |
|---|---|---|---|---|---|
|  | (1) | (2) | (3) | (4) | (5) |
| Forcerank$_{i,t}$ |  |  | 0.28*** | 0.10*** |  |
|  |  |  | (27.20) | (12.82) |  |
| Adj Return$_{i,t}$ | 20.17*** | 62.52*** |  | 60.51*** | 0.01 |
|  | (19.64) | (37.44) |  | (36.10) | (0.75) |
| Adj Return$_{i,t-1}$ | 3.54*** | 4.38*** |  | 4.03*** | -0.01 |
|  | (4.36) | (6.68) |  | (6.23) | (-1.32) |
| Adj Return$_{i,t-2}$ | 3.08*** | 1.72*** |  | 1.41*** | -0.02** |
|  | (3.90) | (3.36) |  | (2.76) | (-2.14) |
| Adj Return$_{i,t-3}$ | 4.02*** | 0.84 |  | 0.44 | -0.02* |
|  | (5.07) | (1.51) |  | (0.79) | (-1.67) |
| Adj Return$_{i,t-4}$ | 3.60*** | 2.07*** |  | 1.71*** | 0.03** |
|  | (4.36) | (3.86) |  | (3.22) | (2.49) |
| Adj Return$_{i,t-5}$ | 3.08*** | 1.68*** |  | 1.37** | 0.01 |
|  | (3.84) | (2.97) |  | (2.45) | (0.88) |
| Adj Return$_{i,t-6}$ | 3.05*** | 0.90 |  | 0.60 | 0.03** |
|  | (3.88) | (1.62) |  | (1.10) | (2.32) |
| Adj Return$_{i,t-7}$ | 2.77*** | 1.75*** |  | 1.47** | -0.02* |
|  | (3.38) | (2.73) |  | (2.35) | (-1.82) |
| Adj Return$_{i,t-8}$ | 2.54*** | 1.69*** |  | 1.44*** | 0.00 |
|  | (3.31) | (3.23) |  | (2.78) | (0.10) |
| Adj Return$_{i,t-9}$ | 1.57** | 0.73 |  | 0.57 | -0.00 |
|  | (1.99) | (1.32) |  | (1.05) | (-0.20) |
| Adj Return$_{i,t-10}$ | 2.08*** | 0.80 |  | 0.60 | -0.01 |
|  | (2.67) | (1.43) |  | (1.07) | (-0.72) |
| Adj Return$_{i,t-11}$ | 0.61 | 1.60*** |  | 1.54*** | -0.02 |
|  | (0.80) | (2.85) |  | (2.78) | (-1.47) |
| Observations | 12,752 | 12,752 | 12,807 | 12,752 | 12,752 |
| R-squared | 0.070 | 0.566 | 0.078 | 0.575 | 0.004 |

**Table IA2. Extrapolation of Past Returns: Humans, ChatGPT, and Realized Returns – 24 Weeks**
This table repeats the regression analysis in Table 2, extending the number of weekly return lags from 12 to 24. Specification (1) uses the consensus Forcerank ranking (ranging from one to ten) as the dependent variable, representing the average ranking of a stock across all participants in a contest, with ten indicating the highest rank and one the lowest. In specifications (2) to (4), the dependent variable is the stock ranking generated by ChatGPT-4 based on stock returns in the past 24 weeks. Specification (5) focuses on one-week-ahead stock returns as the dependent variable. The explanatory variables include lagged returns from week $t-23$ to week $t$. Standard errors are clustered by contest, and *, **, and *** denote statistical significance at the 10%, 5%, and 1% levels, respectively.

|  | Forcerank$_{i,t}$ | | ChatGPT$_{i,t}^{24w}$ | | Return$_{i,t+1}$ |
|---|---|---|---|---|---|
|  | (1) | (2) | (3) | (4) | (5) |
| Forcerank$_{i,t}$ |  |  | 0.23*** | 0.13*** |  |
|  |  |  | (23.63) | (16.19) |  |
| Return$_{i,t}$ | 12.73*** | 33.69*** |  | 32.04*** | 1.05 |
|  | (20.26) | (37.70) |  | (36.76) | (0.73) |
| Return$_{i,t-1}$ | 2.39*** | 9.66*** |  | 9.35*** | -2.67** |
|  | (4.47) | (14.75) |  | (14.49) | (-2.03) |
| Return$_{i,t-2}$ | 2.22*** | 2.53*** |  | 2.25*** | -3.79*** |
|  | (4.25) | (4.58) |  | (4.20) | (-2.69) |
| Return$_{i,t-3}$ | 2.52*** | 2.07*** |  | 1.74*** | -2.64* |
|  | (4.74) | (3.70) |  | (3.22) | (-1.87) |
| Return$_{i,t-4}$ | 2.31*** | 4.34*** |  | 4.04*** | 0.64 |
|  | (4.32) | (6.99) |  | (6.67) | (0.46) |
| Return$_{i,t-5}$ | 2.35*** | 4.41*** |  | 4.11*** | 2.69** |
|  | (4.44) | (7.65) |  | (7.32) | (1.99) |
| Return$_{i,t-6}$ | 1.59*** | 0.59 |  | 0.39 | 5.45*** |
|  | (3.09) | (1.03) |  | (0.70) | (3.63) |
| Return$_{i,t-7}$ | 1.41*** | 1.06* |  | 0.88 | -1.21 |
|  | (2.69) | (1.78) |  | (1.52) | (-0.88) |
| Return$_{i,t-8}$ | 1.32*** | 1.10** |  | 0.93* | -2.11 |
|  | (2.58) | (1.97) |  | (1.73) | (-1.50) |
| Return$_{i,t-9}$ | 0.99** | 0.60 |  | 0.47 | -3.02** |
|  | (1.98) | (1.08) |  | (0.87) | (-2.17) |
| Return$_{i,t-10}$ | 1.83*** | 1.74*** |  | 1.50*** | -0.61 |
|  | (3.71) | (3.07) |  | (2.72) | (-0.43) |
| Return$_{i,t-11}$ | 0.20 | 0.89* |  | 0.86* | -2.00 |
|  | (0.43) | (1.69) |  | (1.68) | (-1.51) |
| Return$_{i,t-12}$ | 0.80* | 0.61 |  | 0.51 | -1.08 |
|  | (1.67) | (1.17) |  | (1.00) | (-0.74) |
| Return$_{i,t-13}$ | 0.99** | 1.05** |  | 0.93* | -0.13 |
|  | (2.20) | (2.00) |  | (1.80) | (-0.09) |
| Return$_{i,t-14}$ | 1.04** | 0.85 |  | 0.71 | 0.71 |
|  | (2.33) | (1.53) |  | (1.32) | (0.55) |
| $Return_{i,t-15}$ | 0.97** | 1.59*** |  | 1.47*** | 1.15 |
|  | (2.03) | (3.05) |  | (2.90) | (0.73) |
| $Return_{i,t-16}$ | 0.92* | -0.56 |  | -0.68 | 3.84*** |
|  | (1.96) | (-0.95) |  | (-1.18) | (2.66) |
| $Return_{i,t-17}$ | -0.39 | -1.05* |  | -1.00* | -0.59 |
|  | (-0.82) | (-1.89) |  | (-1.85) | (-0.46) |
| $Return_{i,t-18}$ | 0.53 | 0.56 |  | 0.49 | -3.54** |
|  | (1.14) | (1.07) |  | (0.95) | (-2.30) |
| $Return_{i,t-19}$ | 1.84*** | -0.29 |  | -0.53 | -3.20** |
|  | (3.94) | (-0.53) |  | (-1.00) | (-2.30) |
| $Return_{i,t-20}$ | 1.12** | 1.55*** |  | 1.40*** | 2.67* |
|  | (2.42) | (2.86) |  | (2.67) | (1.82) |
| $Return_{i,t-21}$ | 0.19 | 0.01 |  | -0.01 | 0.02 |
|  | (0.38) | (0.02) |  | (-0.03) | (0.02) |
| $Return_{i,t-22}$ | 0.77 | 0.21 |  | 0.11 | -0.22 |
|  | (1.57) | (0.36) |  | (0.19) | (-0.15) |
| $Return_{i,t-23}$ | -0.01 | 0.35 |  | 0.35 | 2.31* |
|  | (-0.02) | (0.61) |  | (0.64) | (1.72) |
| Observations | 12,607 | 12,607 | 12,719 | 12,607 | 12,607 |
| R-squared | 0.046 | 0.292 | 0.054 | 0.308 | 0.018 |

**Table IA3. LLM Extrapolation of Price Charts: 24-Week Price Charts**

This table presents the results from linear regressions at the contest-stock-week level, as specified in Eq. (1) in the main text. Forcerank is the average ranking of a stock across all participants in a contest, with ten indicating the highest rank and one the lowest. The dependent variable is the performance ranking generated by ChatGPT-4 based on price charts over the past 24 weeks. The explanatory variables include lagged returns relative to forecast week $t+1$. Standard errors are clustered by contest, and *, **, and *** denote statistical significance at the 10%, 5%, and 1% levels, respectively. The sample period is from February 2016 to December 2017.

|  | ChatGPT Rank$_{i,t}^{24w\ Chart}$ | | |
|---|---|---|---|
|  | (1) | (2) | (3) |
| Forcerank$_{i,t}$ |  | 0.06*** | 0.03*** |
|  |  | (6.40) | (3.44) |
| Return$_{i,t}$ | 6.36*** |  | 5.94*** |
|  | (11.67) |  | (10.81) |
| Return$_{i,t-1}$ | 4.48*** |  | 4.40*** |
|  | (8.43) |  | (8.28) |
| Return$_{i,t-2}$ | 3.76*** |  | 3.69*** |
|  | (7.43) |  | (7.31) |
| Return$_{i,t-3}$ | 3.75*** |  | 3.67*** |
|  | (6.74) |  | (6.63) |
| Return$_{i,t-4}$ | 2.92*** |  | 2.84*** |
|  | (5.70) |  | (5.58) |
| Return$_{i,t-5}$ | 3.36*** |  | 3.28*** |
|  | (6.54) |  | (6.40) |
| Return$_{i,t-6}$ | 2.51*** |  | 2.46*** |
|  | (4.79) |  | (4.71) |
| Return$_{i,t-7}$ | 2.23*** |  | 2.18*** |
|  | (4.38) |  | (4.30) |
| Return$_{i,t-8}$ | 2.36*** |  | 2.32*** |
|  | (4.64) |  | (4.56) |
| Return$_{i,t-9}$ | 1.85*** |  | 1.82*** |
|  | (3.66) |  | (3.61) |
| Return$_{i,t-10}$ | 2.48*** |  | 2.42*** |
|  | (5.29) |  | (5.18) |
| Return$_{i,t-11}$ | 2.10*** |  | 2.10*** |
|  | (4.34) |  | (4.35) |
| Return$_{i,t-12}$ | 1.38*** |  | 1.36*** |
|  | (3.11) |  | (3.07) |
| Return$_{i,t-13}$ | 1.38*** |  | 1.34*** |
|  | (2.81) |  | (2.75) |
| Return$_{i,t-14}$ | 1.92*** |  | 1.89*** |
|  | (3.81) |  | (3.77) |
| Return$_{i,t-15}$ | 0.95** |  | 0.92** |
|  | (2.05) |  | (1.99) |
| Return$_{i,t-16}$ | 1.05** |  | 1.02** |
|  | (2.21) |  | (2.15) |
| Return$_{i,t-17}$ | 0.46 |  | 0.47 |
|  | (0.98) |  | (1.01) |
| Return$_{i,t-18}$ | 0.49 |  | 0.48 |
|  | (1.03) |  | (1.00) |
| Return$_{i,t-19}$ | 1.05** |  | 0.99** |
|  | (2.19) |  | (2.07) |
| Return$_{i,t-20}$ | 0.47 |  | 0.43 |
|  | (0.97) |  | (0.89) |
| Return$_{i,t-21}$ | 0.49 |  | 0.49 |
|  | (0.95) |  | (0.94) |
| Return$_{i,t-23}$ | 0.18 |  | 0.16 |
|  | (0.37) |  | (0.32) |
| Return$_{i,t-23}$ | 0.69 |  | 0.69 |
|  | (1.44) |  | (1.44) |
| Observations | 12,606 | 12,718 | 12,606 |
| R-squared | 0.031 | 0.004 | 0.032 |